\documentclass[aps,pre,twocolumn,10pt]{revtex4-1}
\usepackage{graphicx}
\usepackage{epstopdf}
\usepackage{amsmath}
\usepackage{amssymb}
\usepackage{default}
\usepackage{default}
\newcommand{\be}{\begin{equation}}
\newcommand{\bea}{\begin{eqnarray}}
\newcommand{\bc}{\begin{center}}            
\newcommand{\ee}{\end{equation}}
\newcommand{\eea}{\end{eqnarray}}
\newcommand{\ec}{\end{center}}

\newcommand{\baa}{\begin{eqnarray*}}
\newcommand{\eaa}{\end{eqnarray*}}
\begin{document}
\title{Quantum Otto engine with exchange coupling in the presence
of level degeneracy}
\author{Venu Mehta and Ramandeep S. Johal}
\email{rsjohal@iisermohali.ac.in}
\affiliation{ Department of Physical Sciences, \\ 
Indian Institute of Science Education and Research Mohali,\\  
Sector 81, S.A.S. Nagar, Manauli PO 140306, Punjab, India}
\begin{abstract}
We consider a quasi-static quantum Otto cycle using
two effectively two-level systems with degeneracy in the excited state.
The systems are coupled through isotropic
exchange interaction of strength $J>0$, in the presence
of an external magnetic field $B$ which is varied during the cycle.
We prove the positive work condition, and show that 
level-degeneracy can act as a thermodynamic
resource, so that a larger amount of work can be extracted 
than from the non-degenerate case, both with and without  
coupling. We also derive an upper bound for the efficiency of the cycle.
This bound is the same as derived for a system of coupled spin-1/2 particles
[G. Thomas and R. S. Johal, Phys. Rev. E {\bf 83}, 031135 (2011)] i.e.
without degeneracy, and  
depends only on the control parameters of the Hamiltonian,
but is independent of the level degeneracy and the reservoir temperatures. 
\end{abstract}
\maketitle
\section{Introduction}
\par\indent
The study of thermal machines in the quantum regime, 
is currently under intense investigation. Different quantum working
substances may be employed, such as a three-level maser \cite{scovil}, 
two-level \cite{quan2006,kieu2006,Abe2011} or multi-level quantum particles 
\cite{Johal2008, uzdin2014}
such as spins \cite{george,ivanchenko2015,altintas2015,thomas2016} 
and harmonic oscillators \cite{rezek2006,abah,Lutz2009,agarwal2013}.
This spurt of interest has been due to advances in the 
technology of micro- and nano-machines, as well as theoretical progress on  
the connection between thermodynamics and quantum theory 
\cite{Horodecki2002,Vedral2002,Kosloff2013,Spekkens2013,Kurizki2015}. 
When the size of the  heat engine is reduced, the presence of 
quantum discreteness, degeneracy, and quantum correlations  
pose a natural question: whether the principles of thermodynamics
that were customarily applied to macroscopic objects,
now retain the same form, and if not, what are the consequent modifications? 
Quantum analogs of classical Otto cycle serve as a testbed to 
analyze and illustrate various extensions of thermodynamic
ideas in the quantum domain \cite{zhang2007,wang2009,george,Lutz2012, Li2013,
Thomas2014,
Meystre2014,Stefanotas2014,Wu2014,
hubner2014,Pena2015, ivanchenko2015, Zambrini2015, altintas2015, Jizhou2015,Long2015,
Poletti2015,Campo2016new, Campo2016,Ferdi2016,Leggio2016,Manzano2016,
Salamon2016,Poletti2016,Pekola2016, thomas2016,
Biswas2017,Newman2017}.

Recently, quantum Otto cycle (QOC) in the quasi-static limit, 
with two spin-${1}/{2}$ particles coupled by Heisenberg exchange interaction,
was shown to exhibit an enhanced efficiency \cite{george}. 
Though the efficiency is naturally limited
by Carnot value, a stronger upper bound ($\eta_{\rm ub}$)
was found in the presence of coupling.
This bound depends on the control parameters
in the Hamiltonian, but not on the reservoir temperatures. 
An analysis of local work and 
efficiency shows counter-intuitive effects such as
locally the flow of heat may be in a direction opposite to the 
global temperature gradient \cite{george,Huang2013,Huang2014}.
100\% efficiencies have also been reported using 
negative absolute temperatures of the heat reservoirs \cite{ivanchenko2015}. 

In a related study, the efficiency for 
a system of a spin-$s$ particle coupled to a spin-1/2 particle
\cite{altintas2015} was shown to exceed $\eta_{\rm ub}$,
depending on the spin-value $s$.
An outstanding question is whether  
a bound stricter than the Carnot efficiency may also exist 
for two coupled spins of arbitrary magnitudes $s_1$ and $s_2$.
In this paper, we take a step in this direction
by considering two coupled effectively two-level systems, each with  
a degenerate excited state. Such systems are often studied in 
quantum and atom optics e.g. atoms with V-configuration \cite{Scullybook}, 
where the excited level is very nearly doubly-degenerate.
Within quantum thermodynamics, the role of level-degeneracy
has been explored using finite-time models of thermal machines 
\cite{Johal2008, Kurizki2015n}. 
We show conditions under which work is extractable from such 
a working medium using a quasi-static QOC. 
It will be observed that level-degeneracy can act as a thermodynamic
resource, helping to extract larger work
than the non-degenerate case, both with and without  
coupling. Somewhat surprisingly, we obtain  
 the same upper bound on the efficiency, $\eta_{\rm ub}$, 
 for the degenerate case also. Thus the presence of degeneracy alters
 the extracted work, but not the maximal efficiency of a QOC. 

The paper is organized as follows. In Section II,
we introduce our working substance and the model for QOC.
Various stages of the heat cycle and its efficiency are 
described within a general framework. 
In Section III, we discuss the positive work condition (PWC) and engine's
efficiency for the non-interacting case. In Section IV, we calculate 
the work and efficiency for the interacting case and
PWC is proved within a certain regime of parameter values. 
In Section IV.A, we discuss the upper 
bound to the efficiency which is stricter than the Carnot efficiency.
Section V is on further discussion of our results.  
The final Section VI summarises our main findings.
The computation of eigenvalues of the Hamiltonian, the proofs of PWC
and of the upper bound are given in the appendices.  

\section{Quantum Otto Cycle}
\begin{figure}[ht]
    \includegraphics[width=8cm,height=8cm,keepaspectratio]{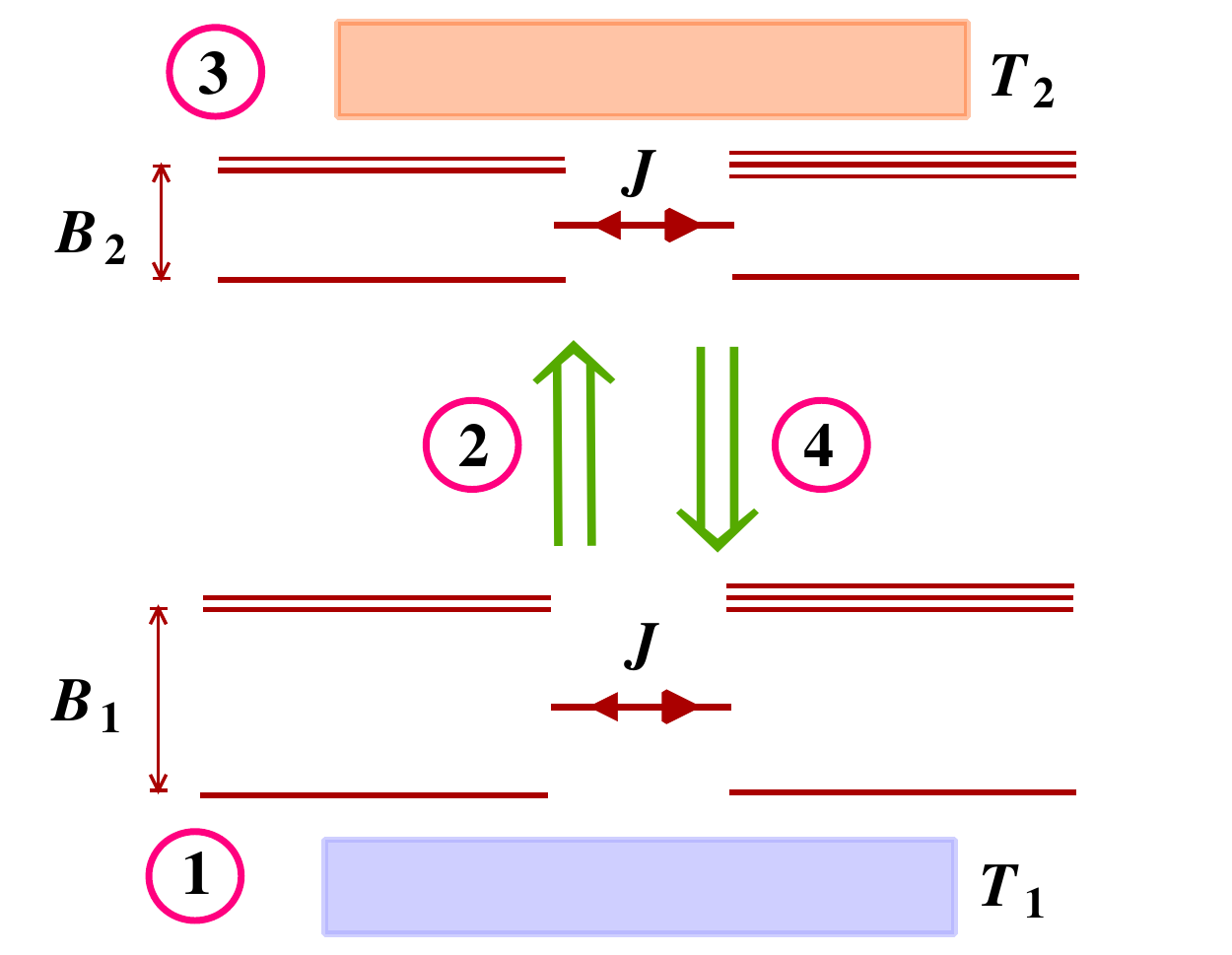}
    \caption{A schematic of QOC between hot ($T_1$)  and 
    cold ($T_2$) heat baths, using working substance as
    two effectively two-level systems
    coupled via exchange interaction of strength $J$. The degeneracy
    of excited levels is $n^{(1)}=2$ and $n^{(2)}=3$. The different
    stages of the cycle are marked with circles.}
    \label{wmedium}
  \end{figure}

The working substance consists of two particles which are
effectively two-level systems such that while the ground state 
is non-degenerate, the excited state of each is $n^{(i)}$-fold 
degenerate, $i=1,2$. The particles are coupled by an Heisenberg
exchange interaction specified below.
The Hamiltonian of the working substance in the first stage 
of the cycle is given by:
\bea
{\cal H}_1  & \equiv & H_{1} + H_{\rm int}, \nonumber \\ 
&=& 2B_1\left( s_{z}^{(1)} \otimes I + I \otimes  s_{z}^{(2)} \right) +  
      8J\overrightarrow{s}^{(1)}.\overrightarrow{s}^{(2)},
\label{Hamiltonian}
\eea
where $H_1$ is the free or local Hamiltonian   
and $H_{\rm int}$ is the interaction Hamiltonian with $J>0$ as the 
anti-ferromagnetic exchange constant.
$\overrightarrow{s}^{(i)} \equiv \{s^{(i)}_{x}, s^{(i)}_{y}, s^{(i)}_{z}| i=1,2\}$, 
are the generalized spin operators for the particles (see Appendix A).
For $n^{(i)} =1$, we have two coupled spin-1/2 particles \cite{george}. 
The various stages of the heat cycle are discussed as follows.
A schematic of the cycle is shown in Fig. \ref{wmedium}.
  
\noindent\textbf{Stage 1}:
The working medium is in thermal equilibrium with a heat reservoir 
at temperature $T_1$.
The density matrix $\rho_1$ for the working substance is given as:
\begin{equation}
\rho_1 =\sum _{i=1}^{N}p_{i}  {\left| \psi _{i}  \right\rangle} {\left\langle \psi _{i}
\right|} 
\label{rho}
\end{equation}  
\noindent where $N=(n^{(1)} +1) (n^{(2)} +1)$ is the number of
energy levels with ${\left| \psi _{i}  \right\rangle}$ as the corresponding 
energy eigenstates. $p_{i} = \exp\left(-E_{i} /T_{1}\right)/Z_{1}$ are 
the occupation probabilities of the microstates with energy  levels ($E_{i} $) and 
$Z_{1}=\sum _{i}\exp \left(-E_{i} /T_{1}\right)$
is the partition function for the system. We set Boltzmann's constant $k=1$.

\noindent \textbf{Stage 2}: 
The system is isolated
from the hot reservoir and the magnetic field is changed from 
$B_{1} $ to $B_{2} $ such that $B_2 < B_1$. So the local Hamiltonian is then given 
as: $H_2 = 2B_2\left( s_{z}^{(1)} \otimes I + I \otimes  s_{z}^{(2)} \right)$.
During this process,  the {\it quantum adiabatic
theorem} is assumed to hold \cite{born1928}, according to which the rate of
the process should 
be slow enough so that the occupation probabilities of the energy 
levels are maintained. Work is performed by the system during this step.

\noindent \textbf{Stage 3}: 
The system is brought in contact with a cold reservoir at temperature 
$T_{2} $ ($< T_{1} $). The energy eigenvalues are $E_{i}^{'}$ 
and the occupation probabilities change from $p_{i}$ 
to $p_{i}'$ corresponding to canonical density matrix $\rho_2$.  
The amount of heat rejected to the cold reservoir is:
\bea
Q_2 & = & {\rm Tr}[{\cal H}_2\Delta \rho ], \nonumber \\
&=& {\rm Tr}[H_2\Delta \rho ] + {\rm Tr}[H_{\rm int} \Delta \rho ],
\label{qc}
\eea
where $\Delta \rho = (\rho_1 - \rho_2)$. 
The first term above can be identified with the amount
of heat {\it locally} exchanged by the working substance,
while the second term is solely due to the interaction. 

\noindent \textbf{Stage 4}: 
The system is detached from the cold reservoir and the magnetic field is 
increased from $B_{2} $ to $B_{1}$ with occupation
probabilities remaining unchanged at $p_{i}' $ and 
energy eigenvalues change from $E_{i}^{'} $ to $E_{i} $.
Work is performed on the system.

\noindent Finally the system is attached to the hot reservoir
again. Thus the system returns to the initial state, completing the 
heat cycle. 
The amount of heat absorbed from the hot reservoir in one cycle is:
\bea
Q_1 & = &  {\rm Tr}[{\cal H}_1\Delta \rho ], \nonumber \\
&=& {\rm Tr}[H_1\Delta \rho ] + {\rm Tr}[H_{\rm int} \Delta \rho ],
\label{qh}
\eea
The work performed per cycle is:
\be
W = Q_1 - Q_2 =  {\rm Tr}[(H_1-H_2) \Delta \rho ]. 
\label{work}
\ee
The efficiency $\eta = W/Q_h$ can be written as:
\bea
\eta &=& \left( 1 - \frac{{\rm Tr}[H_2 \Delta \rho ]}{{\rm Tr}[H_1\Delta \rho ]}
\right)
\left( 1+ \frac{{\rm Tr}[H_{\rm int} \Delta \rho ]}
{{\rm Tr}[H_1\Delta \rho ]}  \right)^{-1}, \nonumber \\
   &=&\eta_{\rm loc} \left( 1+ \frac{{\rm Tr}[H_{\rm int} \Delta \rho ]}
{{\rm Tr}[H_1\Delta \rho ]}  \right)^{-1}.
\label{ef24}
%
%
\eea
The first factor, $\eta_{\rm loc}$, based on the amounts
of heat exchanged locally with the system, can be regarded as the  
efficiency in a local sense. 
This factor also becomes equal to the efficiency
of the engine when the interactions are absent.

In this paper, we are considering the Hamiltonian for which 
we can write $H_1 \equiv 2 B_1 h_0$, $H_2 \equiv 2 B_2 h_0$, i.e. 
the free Hamiltonian is proportional
to the control parameter $B$, and $H_{\rm int} \equiv 8 J h_{\rm int}$, 
where $J$ is held fixed. 
Then we have, $W = 2(B_1-B_2) {\rm Tr}[h_0 \Delta \rho]$.
In this case, 
\be
\eta_{\rm loc} = 1 - \frac{B_2}{B_1} = \eta_{0},
\ee
where $\eta_0$ is the efficiency in the absence of the 
interaction \cite{Comment2}.

Further, as locally the system works like an engine,
so ${\rm Tr}[H_1\Delta \rho ] >0$. From Eq. (\ref{ef24}),
if ${\rm Tr}[H_{\rm int} \Delta \rho ] >0$, then
the efficiency is less than the uncoupled case.
In this paper, we are interested in the situation
${\rm Tr}[H_{\rm int} \Delta \rho ] <0$,  
whereby the  efficiency becomes greater than $\eta_{\rm loc}$, or $\eta_0$
\cite{Comment}. For convenience,  we
write Eq. (\ref{ef24}) as follows:
\be
\eta = \eta_{\rm 0} \left( 1 - \frac{4 J Y}{B_1 X} \right)^{-1},
\label{etgen}
\ee
where  $X = {\rm Tr}[h_0\Delta \rho ] >0$  and
$Y = -{\rm Tr}[h_{\rm int} \Delta \rho]>0$.
Now, under certain conditions, it can be proved that 
$X > Y$. This ensures that an upper bound exists for the efficiency,
given as:
\be
\eta \le \eta_{0} \left( 1 - \frac{4J}{B_1} \right)^{-1}.
\ee
This bound 
is independent of the reservoir temperatures, and is 
tighter than the Carnot limit, within a certain range of parameter values. 
In Ref. \cite{george}, the above bound 
was proved for the working substance of two 
coupled spin-1/2 systems. In the following, 
we make a detailed analysis of work and efficiency
in a QOC with our chosen working substance. 
\section{The uncoupled model}
\label{qheuncoupled}
We first summarise the cycle in the non-interacting case. 
The working substance consists of 
two non-interacting particles in
the presence of an externally applied magnetic field $B$ 
along $z$- axis such that each particle effectively has two energy levels with 
a non-degenerate ground state, but the excited state may have an arbitrary   
degeneracy $n^{(1)}$ and $n^{(2)}$, respectively. 
The Hamiltonian of this system is $H_1$.
Since the particles are non-interacting, each particle undergoes an independent
heat cycle.  For a particle with 
degeneracy $n^{(i)}$, the probabilities at Stage 1 are:
$ p_{1} = e^{B_{1} /T_{1} } /z_{1}$ and $p_{2}= n^{(i)} e^{-B_{1} /T_{1}}/z_{1}$,
where $z_{1} = e^{B_{1} /T_{1} }+ n^{(i)} e^{-B_{1} /T_{1} }$. 
Similarly, the occupation probabilities at Stage 3 are:
${p_{1}' = e^{B_{2} /T_{2} }/z_{2} }$ and
${p_{2}' = n^{(i)} e^{-B_{2} /T_{2}}/z_{2} }$, 
where  
$z_{2} = e^{B_{2} /T_{2} }+ n^{(i)} e^{-B_{2} /T_{2} }$. 
Then the heat absorbed from the hot reservoir is: 
$q_{1} =2B_{1} \left(p_{2}^{} -p_{2}' \right)$,
and the heat rejected to cold reservoir is:
$q_{2} = 2 B_{2} \left(p_{2}^{} -p_{2}' \right)$.
\noindent The work extracted per particle during the heat cycle is
\begin{equation}
w=q_{1} - q_{2} =2(B_{1} -B_{2}) \left(p_{2}^{} -p_{2}' \right).
\label{work un}
\end{equation}
Since $B_{1} > B_{2}$, so for a positive work
condition (PWC), we must have $p_{2} > p_{2}'$, which means
\begin{equation}
\frac{B_{2}}{T_{2}} > \frac{B_{1}}{T_{1}}.
\label{condition un}
\end{equation}
As shown in Appendix B, the 
extracted work from such a system with $n^{(i)}$-fold degeneracy,
is bounded from above by the work from $n^{(i)}$ two-level 
systems (without degeneracy). Thus degeneracy can 
act as a thermodynamic resource.

The efficiency $\eta _{0} ={w}/{q_{1} }$, is 
given by $1- {B_{2} }/{B_{1}}$. 
Due to (\ref{condition un}), the engine efficiency satisfies:
\begin{equation}
\eta _{0} = 1-\frac{B_{2} }{B_{1} } < 1-\frac{T_{2} }{T_{1} }=\eta_{C}, 
\label{eff un}
\end{equation}
where $\eta_{C}$ is the Carnot efficiency.

  \section{The coupled model}
\noindent We now switch on the coupling between the particles.
The Hamiltonian of this system for $B=B_1$ is as given in Eq. \ref{Hamiltonian}.  
The eigenvalues of the Hamiltonian (see Appendix A for an example
with particular values of degeneracies),
ordered in increasing energy, are as follows:
\[
\begin{array}{l l} {\rm Energy}\; (E_{i}) & {\rm Degeneracy}\; (g_{i}) \\
-2B_1 + 2J & 1  \\
-6J & 1 \\
-2J & n^{(1)} + n^{(2)} - 2 \equiv m \\
+2J  & 1 \\
+2B_1 + 2J & n^{(1)} n^{(2)} \equiv n 
\end{array}
\]
We take parameter $J$ to satisfy $B>4J$, so that 
$-2B+2J$ is the ground state.
In this paper, we study the effect of coupling strength $J$
on thermodynamic properties of QOC, for given values of 
$B_1, B_2, T_1$ and $T_2$. 
Now the occupation probabilities for the above energy levels can be calculated as:
$P_{i}= {g_{i} e^{-E_{i} /T }}/{\sum _{i} g_{i} e^{-E_{i} / T }}$. 
We denote the probabilities at {Stage 1} 
by $P_{i}$ with $B=B_1$ and temperature $T_1$, and that at {Stage 3} by 
$P_{i}^{'}$ with $B=B_{2}$ and temperature $T_{2}$.
Note that for the simplest case of two spin-1/2 particles
where $n^{(1)} = n^{(2)} = 1$, we have only four distinct energy 
levels---without degeneracy. 
Further, the parameter values satisfy Eq. (\ref{condition un}).

Now, the heat absorbed by the working medium during 
Stage 1, and the heat rejected to the sink during Stage 3 
respectively is:
\begin{equation}
Q_1 = 2B_1 X - 8 J Y, \quad Q_2 = 2 B_2 X - 8JY,
\label{heat abs HT}
\end{equation}
 where 
 \bea
 X &=& P_{1}^{'} -P_{1} + P_{5} -P_{5}^{'},  \label{X} \\
Y &=& P_{2}^{} -P_{2}^{'} + \frac{1}{2}(P_{3}^{} -P_{3}^{'} ).
\label{Y}
\eea
\noindent The work performed in one cycle is:
\begin{equation}
W=Q_{1} - Q_{2} =2(B_{1} -B_{2}) X.
\label{work HT}
\end{equation}
Then as shown in Appendix B, the PWC
i.e. $W>0$ holds as long as $J \leq J_c$, where
\be
J_c = \frac{1}{4} {\left(\frac{1}{T_{2} } -\frac{1}{T_{1} } \right)^{-1}}
\left({\frac{B_{2} }{T_{2} } -\frac{B_{1} }{T_{1} } } \right).
\label{jc}
\ee
%
 \begin{figure}[ht]
  \centering
    \includegraphics[width=8cm,height=8cm,keepaspectratio]{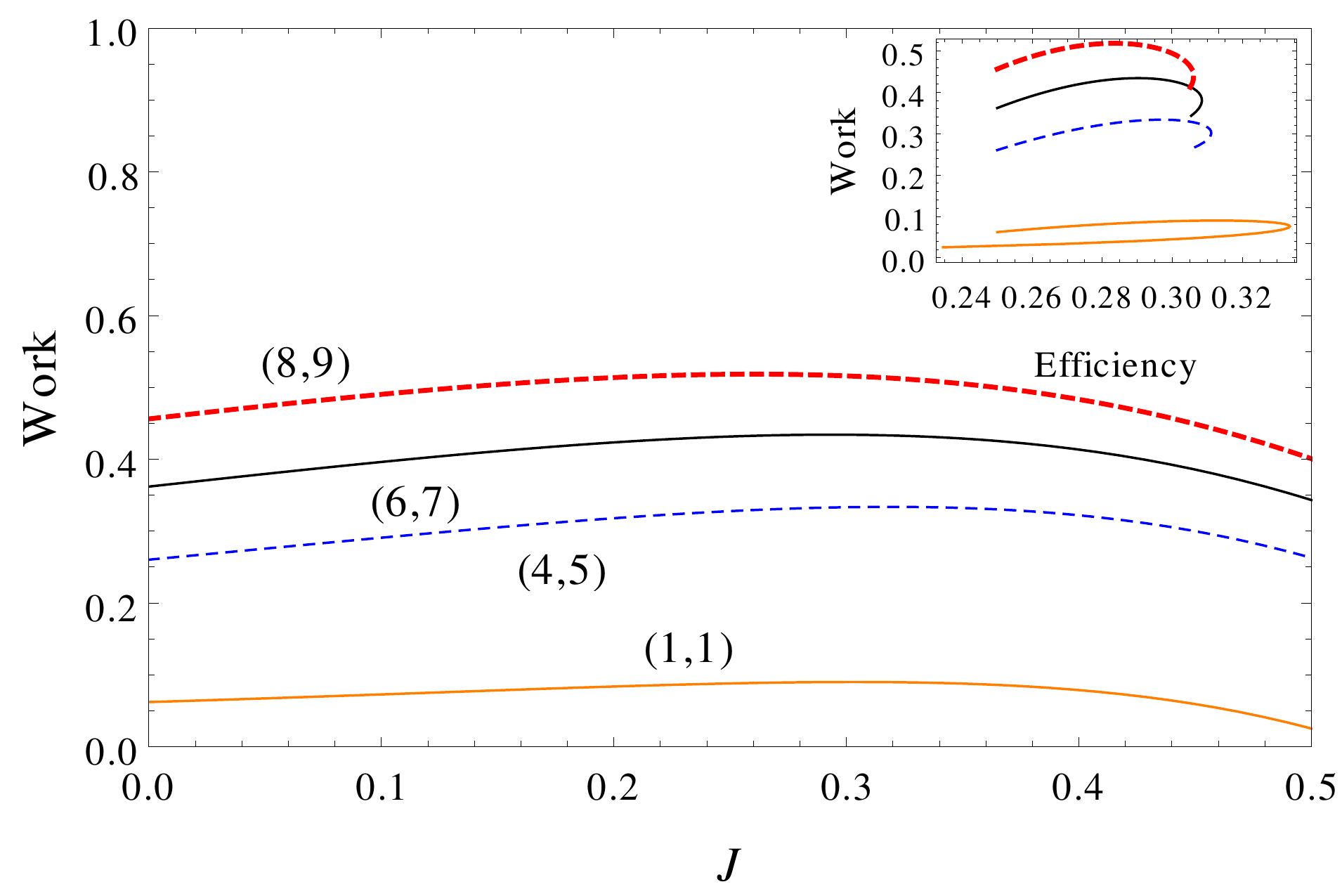}
    \caption{The work extracted from the engine as a
function of the coupling strength $J$ for different $(n^{(1)},n^{(2)})$. 
The parameter values are $B_{1}$=4, $B_{2}$=3,
$T_{1}=2$ and $T_{2}=1$. The inset shows the corresponding parametric plots 
of work versus efficiency as function of parameter $J \in [0,J_c]$, traced 
in a clockwise fashion. The loop-shaped
curves show that work and efficiency achieve their maximum values
at different values of $J$.}
    \label{wvsj}
  \end{figure}
%
In other words, for given values $B_1, B_2, T_1$ and  $T_2$,
satisfying Eq. (\ref{condition un}), a sufficient criterion
for $W>0$ in the coupled case is that $J \le J_c$.
In Fig. \ref{wvsj}, it is shown that degeneracy of the 
excited level leads to a higher work extraction.

\subsection{Efficiency and its upper bound}
\begin{figure}[ht]
  \centering
    \includegraphics[width=8cm,height=8cm,keepaspectratio]{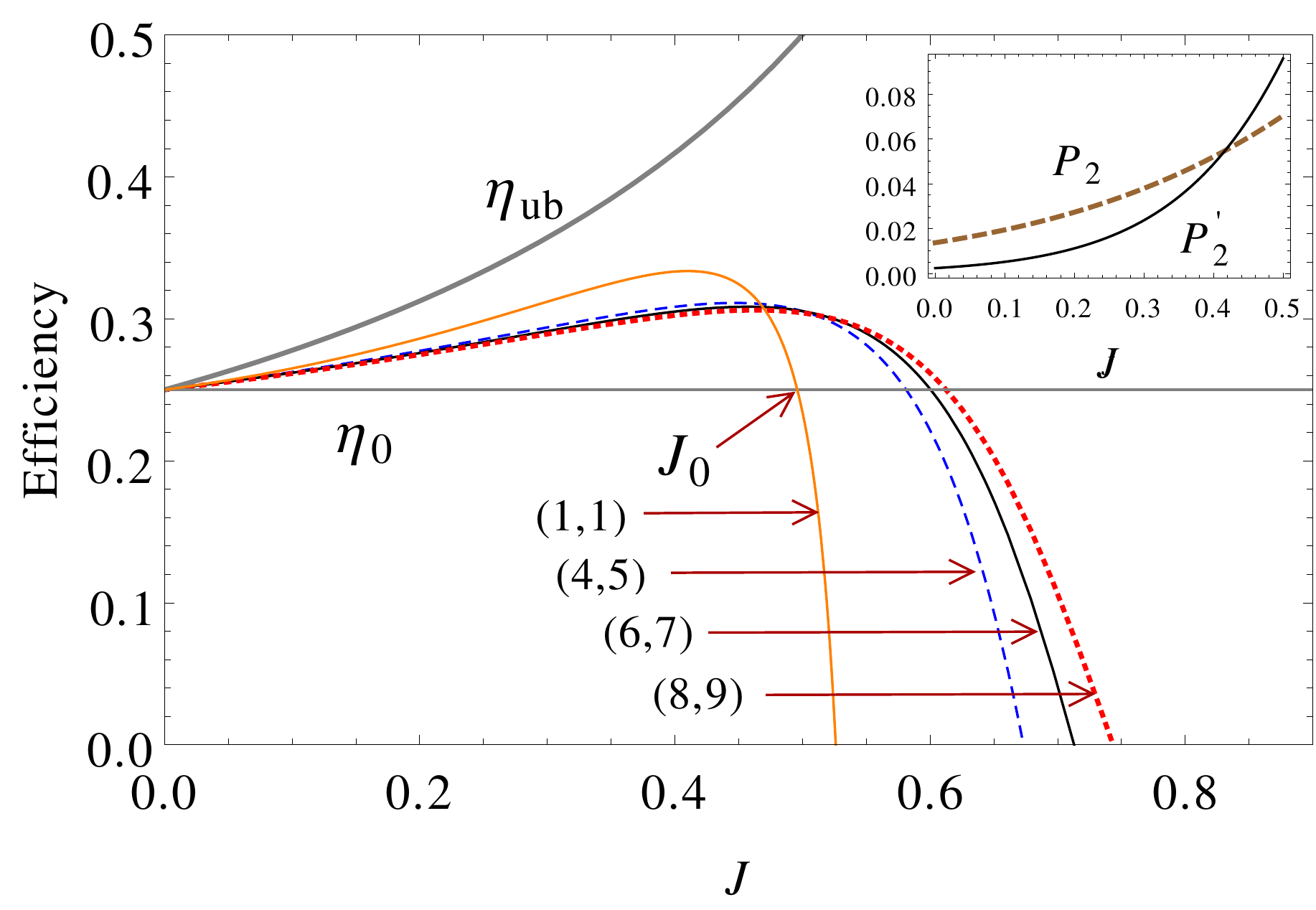}
    \caption{The variation of efficiency of the engine  as a
function of the coupling strength $J$, for different $(n^{(1)},n^{(2)})$. 
The parameter values are $B_{1}$=4, $B_{2}$=3,
$T_{1}=2$ and $T_{2}=1$. $J_0$ marks the value of $J$ where efficiency
for (1,1) case (without degeneracy)
cuts the uncoupled efficiency line, $\eta_0=0.25$. Upper bound $\eta_{\rm ub}$ is also shown.
The inset shows the regimes $P_2 > P_2'$ and $P_2' > P_2$ vs. $J$ for (4,5) case; the 
curves for probabilites cross each other at a value of $J<J_c=0.5$.}
    \label{effvsj}
  \end{figure}

Now, from Eq. (\ref{etgen}), if $J=0$, 
the efficiency is equal to $\eta_{0}$. It is observed that $Y(J)$ as 
function of $J$ shows a non-monotonic
behavior, so that $\eta$ first increases with $J$, but 
then decreases, and beyond a certain $J(\equiv J_m)$, becomes less 
than $\eta_{0}$, (see  Fig. \ref{effvsj}).
Within the general discussion of Section II, we can say
that the efficiency is equal to $\eta_{0}$ when 
${\rm Tr}[H_{\rm int} \Delta \rho ] = 0$. This happens
in the case of no interactions, $J=0$, but may
have a solution for some non-zero interaction strength.
At the latter point, we have then the condition
${\rm Tr}[H_{\rm int} \rho_1] = {\rm Tr}[H_{\rm int} \rho_2 ]$,
i.e. the average interaction energy is not altered on 
changing the equilibrium state of the working substance
from state $\rho_1$ to $\rho_2$.

So $J_m$ is determined by $Y (J_m) = 0$. 
This implies, from Eq. (\ref{Y}),  
$P_3 + 2 P_2 = P_3' + 2 P_2'$. Expressing $P_3$ ($P_3'$) in terms of
$P_2$ ($P_2'$), this condition is rewritten as:
\be
( m e^{-4 J_m/T_1} + 1) P_2 = (m e^{-4 J_m/T_2} + 1) P_2'.
\label{y10}
\ee
First, we take  the case
of two spin-1/2 particles ($m=0$). 
Then $P_2 = P_2'$ is the condition for $Y(J_0)=0$. 
In this case, we can explicitly solve for $J_0$:
\be
J_0 = 
\frac{1}{8}\left(\frac{1}{T_{2} } -\frac{1}{T_{1} } \right)^{-1} \ln \left(
\frac{1+ e^{2B_2/T_2} + e^{-2B_2/T_2} }{1+ e^{2B_1/T_1} + e^{-2B_1/T_1} } \right).
\label{j0half}
\ee
For the above case, it can be shown that $J_0 \le J_c$, 
the equality being approached for low temperatures.
However, in the case of degeneracy ($m\ne 0$), 
it is not possible to find a closed expression for $J_m$.
Numeric evidence shows that degeneracy ($m\ne 0$) may lead to 
an extended regime of enhanced efficiency, beyond $J=J_c$ values (see Fig. 3).
  
For the uncoupled or $J=0$ case, PWC leads to condition $p_2 > p_2'$.
As $J$ becomes non-zero, we have $P_2 > P_2'$ up to a
certain value of $J$, after which the opposite condition i.e. 
$P_2' > P_2$, holds.  
It is convenient to do further analysis in the following
regimes, separately:
(i) $P_2 > P_2'$, (ii) $P_2' > P_2$ (see inset of Fig. \ref{effvsj}).

(i) $P_2 > P_2'$ {\it Regime}:
It can be shown (Appendix C) that 
the condition $P_2 > P_2'$ implies the following relations between 
the occupation probabilities:                                                                        
\begin{equation}
P_{1} <P_{1}^{'} \quad P_{3} >P_{3}^{'}, \quad P_{4} >P_{4}^{'}, \quad P_{5}>P_{5}^{'}.
\label{occ prob rela}
\end{equation}
%

PWC has already been proved, but here due to
$P_{5} >P_{5}^{'}$ and $P_{1} <P_{1}^{'}$, it is clear that 
 $X>0$, or $W>0$. 
From Eqs. (\ref{heat abs HT}) and (\ref{work HT}), 
the efficiency of the engine in the general case is:
\begin{equation}
\eta  =
\frac{\left(B_{1} -B_{2} \right)X}{B_{1} X - 4 J Y} =
{\eta _{0} }{ \left( 1 - \frac{4JY}{B_{1}X } \right)^{-1} }.
\label{etap}
\end{equation}
Further, in this regime, due to Eq. (\ref{occ prob rela}), we also have  
$Y>0$, which implies that the efficiency, Eq. (\ref{etap}),
is higher than uncoupled case: $\eta > \eta_0$.
Further, it can be proved (Appendix C) that $X>Y$. 
Therefore, we can write:
\[\eta \leq
{\eta _{0} } \left({1-\frac{4J}{B_{1}} }\right)^{-1} \equiv \eta_{\rm ub}.\]
   
(ii) $P_2' > P_2$ {\it Regime}:
Now, we have seen that $Y>0$ is a necessary condition for 
$\eta > \eta_0$. So from Eq. (\ref{Y}), when $P_2' > P_2$,
then in order to have an enhancement of efficiency,
we must have $P_3 > P_3'$. Then we also obtain                              
$P_{4} >P_{4}^{'}$ and $P_{5} >P_{5}^{'}$. Note that this still leaves
the inequality between $P_1$ and $P_1'$ undetermined. However,
it can be shown here also that $X>Y$ (Appendix C) 
and so the same upper bound on efficiency 
holds in this regime also. 

Finally, it can be directly seen that 
the upper bound $\eta_{\rm ub}$ derived above 
is less than the Carnot limit $\eta_C$, 
as long as $J < J_c$, where $\eta_{\rm ub} = \eta_{C}$,
for $J=J_c$.

\section{Discussion}
\begin{figure}[ht]
  \centering
    \includegraphics[width=8cm,height=8cm,keepaspectratio]{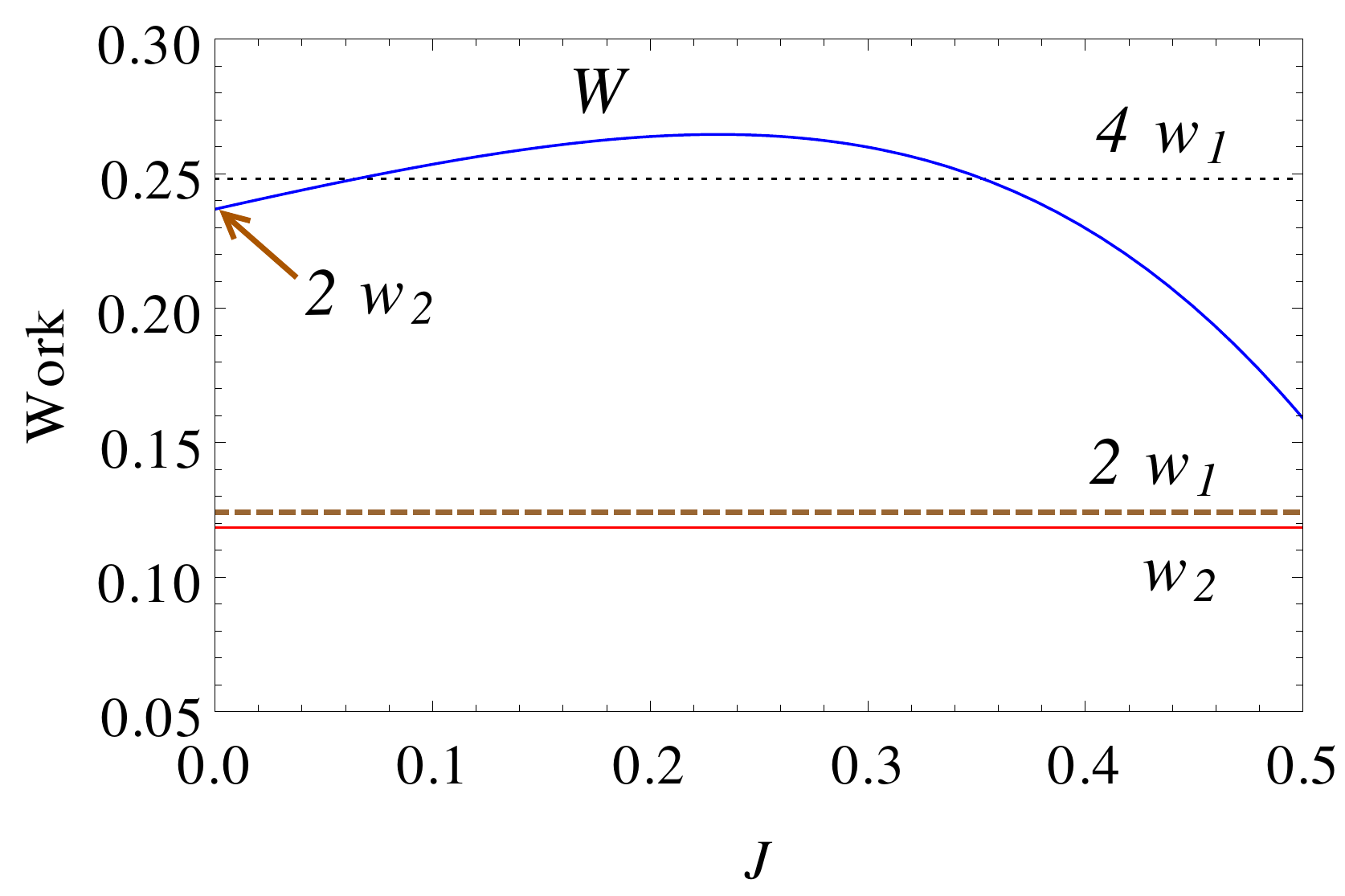}
    \caption{The  extracted  work ($W$) in a QOC with two interacting V-configuration
    systems ($n^{(1)} = n^{(2)}=2$)  as a
function of  $J$, compared with 
two such non-interacting systems ($2 w_2$), as well as 
four non-interacting two-level systems ($4 w_1$). The lines showing 
$w_2 < 2w_1$ illustrate that $w_n$ is bounded by $n w_1$ (see Appendix B).
The parameter values are set at $B_{1}$=4, $B_{2}$=3,
$T_{1}=2.5$ and $T_{2}=1.25$.}
    \label{wvsj2}
  \end{figure}

In an earlier study \cite{Johal2008},
and more recently \cite{Kurizki2015n}, level degeneracy has been
shown to act as a thermodynamic resource. In Ref. \cite{Johal2008},
a two-step, finite time cycle was considered using two multilevel systems,
each in contact with its respective reservoir. The working 
medium is regarded as isolated from the reservoirs during the work extraction step. 
However, in Ref. \cite{Kurizki2015n}, the working medium--- 
in the form of a V-configuration system, is continuously
coupled to both the reservoirs during the cycle.  In the present work,
using a quasi-static, four-step Otto cycle on such systems,  it is 
 shown that degeneracy can act as a thermodynamic resource
 and the work ($w_n$) extracted from a two-level system with $n$-fold degeneracy
 in the excited level, is bounded from above by the work ($w_1$) obtained
 from $n$ two-level systems, i.e. $w_n  \leq n w_1$ (see Appendix B).  
 For the special case of a V-configuration system, it implies 
 $w_2 \leq 2 w_1$. This result 
 may be compared to an analogous result on the power extracted
 in case of a finite-time model \cite{Kurizki2015n} using the same system. 
 By the same token, the work per cycle from two non-interacting V-configuration systems 
 would be bounded by the work from four two-level systems. However, 
 as we show by numerical evidence in Fig. \ref{wvsj2}, the presence of the exchange coupling
 between two V-configuration systems may lead to 
 a higher amount of extracted  work than from four two-level systems.
\section{Summary}
We analyzed quantum Otto engine between two heat reservoirs, using two  
effectively two-level particles with degeneracy in the excited state.
In our model, the particles are coupled by 
isotropic exchange interaction in the presence 
of an external magnetic field $B$, varied during the cycle.
We could show that for given values of $B_1, B_2, T_1$ and  $T_2$, satisfying
${B_{2}}/{T_{2}} > {B_{1}}/{T_{1}}$ (PWC for the uncoupled case),
a sufficient criterion
for  $W>0$ in the coupled case is that $J \le J_c$ where $J_c$ 
is a certain critical value (independent of degeneracy).
It is shown that coupling can lead to
enhancement in both the extracted work and the efficiency.
An upper bound to the efficiency is derived in the regime 
$J< J_c$, which also implies that the bound itself is limited
by Carnot value. Interestingly, the bound is independent of
the degeneracy of the levels and is thus equal to the one found
for two coupled spin-1/2 particles \cite{george}. 
On the other hand, we find that degeneracy can act as thermodynamic
resource, helping to extract a larger amount of work, with or
without coupling. 
Further, numerical evidence shows 
that the degeneracy ($m\ne 0$) leads to an enhancement of work
and efficiency even in the strong
coupling regime ($J>J_c$). A corresponding analysis in this
regime could shed more light on the role of degeneracy in
the performance of the idealized cycle.

\section*{Acknowledgements}
Venu Mehta acknowledges financial support in the form 
of Senior Research Fellowship, from IISER Mohali.

\appendix
\section{Generalized spin operators and the Hamiltonian}
In the main text, we assumed arbitrary values for the degeneracies
of excited states of the two particles. In the following, 
we calculate the 
eigenvalues of the Hamiltonian for a particular example: when
one particle has a non-degenerate ground state and a 
two-fold degenerate excited state ($n^{(1)}=2$), while 
the other particle has four energy levels with one 
ground state and the excited state as three-fold degenerate ($n^{(2)}=3$).
Therefore, we have $m= n^{(1)} + n^{(2)}-2 = 3$, and $n=n^{(1)}n^{(2)}= 6$.
For the first particle, the operator $s_{z}^{\left(1\right)} $
is written as:
\[s_{z}^{\left(1\right)} =\frac{1}{2}\left(\begin{array}{ccr} {1} & {0} & {0} \\
 {0} & {1} & {0} \\ {0} & {0} & {-1} \end{array}\right), \]
 where Planck's constant has been set to unity.
The normalized eigenstates of $s_{z}^{\left(1\right)}$: 
\noindent                                                                       
\[\left(\begin{array}{l} {1} \\ {0} \\ {0} \end{array}\right),
\left(\begin{array}{l} {0} \\ {1} \\ {0} \end{array}\right),
\left(\begin{array}{l} {0} \\ {0} \\ {1} \end{array}\right)
\]
denoted as $\left| +  \right\rangle,\left| 0  \right\rangle,\left|
-  \right\rangle $, corresponding to eigenvalues $1/2,1/2,-1/2$
respectively.

We define the raising operator as :
$s_{+}^{\left(1\right)} ={| \alpha\rangle}\langle -|$,
where $| \alpha \rangle = ({|+\rangle}+{|0\rangle})/{\sqrt{2} }$.
This implies that the action of $s_{+}^{\left(1\right)}$
on $|-\rangle$ takes the particle into a superposition
of the two (degenerate) excited states, with equal amplitudes ($1\sqrt{2}$). 
In matrix form, we obtain
\[s_{+}^{\left(1\right)} =\frac{1}{\sqrt{2}}\left(\begin{array}{ccc} {0} & {0} & {1 } 
\\ {0} & {0} & {1 } \\ {0} & {0} & {0} \end{array}\right).\] 
The lowering operator defined as $s_{-}^{(1)} = \left(s_{+}^{(1)} \right)^{\dag }$,
is given by:
\[s_{-}^{(1)} =\frac{1}{\sqrt{2}}
\left(\begin{array}{ccc} {0} & {0} & {0} \\ {0} & {0} & {0} \\
{1}  & {1} & {0} \end{array}\right).\] 
The operators $s_{x}^{\left(1\right)}=
(s_{+}^{(1)} + s_{-}^{(1)})/2$ and
$s_{y}^{\left(1\right)}=({s_{+}^{(1)} -
s_{-}^{(1)} })/{2i}$ are given as:
\[s_{x}^{\left(1\right)} =\frac{1}{2\sqrt{2}}\left(\begin{array}{ccc} {0} & {0} & {1 }
 \\ {0} & {0} & {1} \\ {1 } & {1} & {0} \end{array}\right), \quad
s_{y}^{\left(1\right)} = \frac{1}{2\sqrt{2} i}\left(\begin{array}{rrr} {0} & {0} & {1} \\ 
{0} & {0} & {1} \\ {-1} & {-1} & {0} \end{array}\right).\]
The particle operators for the other particle 
with one ground state and the excited state is 3-fold  degenerate,
are similarly calculated and given as follows.
\[s_{z}^{\left(2\right)} =\frac{1}{2}
\left(\begin{array}{cccc} {1} & {0} & {0} & {0} \\ 
{0} & {1} & {0} & {0} \\ {0} & {0} & {1} & {0} \\ 
{0} & {0} & {0} & {-1} \end{array}\right),\] 
and
\[s_{x}^{\left(2\right)} =
\frac{1}{2\sqrt{3}}\left(\begin{array}{cccc} {0} & {0} & {0} & {1} \\ {0} & {0} & {0} & {1 }
 \\ {0} & {0} & {0} & {1} \\ {1 } & {1 } & {1 } & {0} \end{array}\right),                    
s_{y}^{\left(2\right)} =
\frac{1}{2\sqrt{3}i}\left(\begin{array}{rrrc} {0} & {0} & {0} & {1} \\ {0} & {0} & {0} & {1} 
\\ {0} & {0} & {0} & {1} \\ {-1} & {-1} & {-1} & {0} \end{array}\right).\]
Then the eigenvalues of the Hamiltonian 
\bea
 {\cal H}_1 & = & 2B_1( s_{z}^{(1)} \otimes I + I \otimes  s_{z}^{(2)} ) \nonumber \\
  & & +  
      8J (s_{x}^{(1)} \otimes s_{x}^{(2)}+ 
      s_{y}^{(1)}\otimes s_{y}^{(2)}+s_{z}^{(1)}\otimes s_{z}^{(2)})
      \nonumber,
\eea
are computed to be:
\[
\begin{array}{r c} {\rm Energy}\; (E_i) & {\rm Degeneracy}\; (g_i) \\
-2B_1 + 2J & 1  \\
-6J & 1 \\
-2J & 3 \\
+2J  & 1 \\
2B_1 + 2J & 6 
\end{array}
\]  
\section{Degeneracy as a Thermodynamic Resource}
The work extracted from the $n$-fold degenerate system is given by
\be 
w_n = 2n(B_1-B_2)\left( \frac{1}{e^{2B_1/T_1} + n}  - \frac{1}{e^{2B_2/T_2} + n}    \right).
\ee
Let $w_1$ represent the work extracted from a non-degenerate two-level system.
In the following, we denote $e^{2B_1/T_1} = x_1$ and $e^{2B_2/T_2} = x_2$. 
We can write:
\bea
w_n - n w_1 &=& 2(B_1-B_2)C^{-1} (x_2 - x_1) \times \nonumber \\
&& \left[ n(1-n^2) + n(1-n)(x_1+x_2)     \right],
\eea
where $C = (x_1+1)(x_1+n)(x_2+1)(x_2+n)>0$. Since $B_1> B_2$ and $x_2 > x_1$, we note
that 
\be 
w_n - n w_1 \leq 0.
\ee
Thus the work per cycle from the $n$-fold degenerate system is bounded from above
by the work from a working medium of $n$ non-degenerate, two-level systems.
\section{Positive Work Condition (PWC)}
\noindent The work extracted per cycle in the coupled case is given by:
\[W=2(B_{1} -B_{2} )\left(P_{1}^{'} -P_{1} 
 + P_{5} -P_{5}^{'} \right).\] 
Since $B_{1} >B_{2} $ so we need to show $P_{1}^{'} -P_{1} + P_{5} -P_{5}^{'} >0$,  
or  $P_{1}^{'} -P_{5}^{'} > P_{1} -P_{5}$. This implies 
\[
Z_{2}^{-1} {(e^{2B_{2} /T_{2} } -n e^{-2B_{2} /T_{2} })}
>Z_{1}^{-1} {(e^{2B_{1} /T_{1} } - n  
e^{-2B_{1} /T_{1} })}. \] 
On cross multiplying and rearranging terms, the above inequality can
be written as:
\[\begin{array}{l} (e^{2B_{2} /T_{2} } -e^{2B_{1} /T_{1}} ) \\
+2n (e^{-2B_{1} /T_{1} +2B_{2} /T_{2} } -e^{2B_{1} /T_{1} -2B_{2} /T_{2} }) \\
+ n (e^{-2B_{1} /T_{1} } -e^{-2B_{2} /T_{2} }) \\ 
+ m n (e^{-2B_{1} /T_{1} +4J/T_{2} } -e^{-2B_{2} /T_{2} +4J/T_{1} } ) \\
+n (e^{-2B_{1} /T_{1} +8J/T_{2} } -e^{-2B_{2} /T_{2} +8J/T_{1} }) \\ 
+ {m (e^{2B_{2} /T_{2} +4J/T_{1} } -e^{2B_{1} /T_{1} +4J/T_{2} })} \\ 
+ (e^{2B_{2} /T_{2} +8J/T_{1} } -e^{2B_{1} /T_{1} +8J/T_{2} } )>0. 
\end{array}\] 
We can check that for $T_1>T_2$, every term enclosed by parentheses is positive,
if we apply the following conditions:
\begin{itemize}
\item $\frac{B_{2} }{T_{2} } > \frac{B_{1} }{T_{1} }$,
\item $J \le J_c = \frac{1}{4}
\left(\frac{1}{T_{2}} -\frac{1}{T_{1}} \right)^{-1}
\left(\frac{B_{2} }{T_{2} } -\frac{B_{1} }{T_{1} }\right)$.
\end{itemize}
The above set of conditions are sufficient to ensure the PWC.
\section{Proof of upper bound for the efficiency}
\subsection{$P_{2}^{} > P_{2}^{'}$}
We assume the inequality: $P_{2}^{} > P_{2}^{'}$. 
From the definitions of the probabilites, we can write
$P_{3}  = m e^{-4J/T_{1} } P_{2}$,
and $P_{3}^{'} = m e^{-4J/T_{2} } P_{2}^{'}$.
For $T_{1} >T_{2}$ and $J>0$, this implies  
$P_{3} >P_{3}^{'}$. Further, due to
$P_{4} = e^{-8J/T_{1} } P_{2}$ and  
$P_{4}^{'} =e^{-8J/T_{2} } P_{2}^{'}$,
it follows that $P_{4} >P_{4}^{'}$. Finally,  
as $P_{5} = n e^{-(2B_{1} /T_{1} +8J/T_{1}) } P_{2}$, and
$P_{5}^{'} =n e^{-(2B_{2} /T_{2} +8J/T_{2} )} P_{2}^{'}$,
so  $P_{5} >P_{5}^{'}$. 
Using the normalization condition on probabilites,
$\sum_{i=1}^{5} P_i = \sum_{i=1}^{5} P_i'=1$, 
we finally obtain: $P_{1} <P_{1}^{'}$.
Thus if $P_2 > P_2'$ is true, then definite inequalities
exist between all the primed and unprimed probabilites:
\be
P_{1}^{'} > P_{1}^{}, \quad P_{3}^{} > P_{3}^{'},
\quad P_{4}^{} > P_{4}^{'}, \quad P_{5}^{} > P_{5}^{'}.
\label{ppp}
\ee
From the above relations, we see that $Y>0$, i.e.
the efficiency is higher than $\eta_0$. 
Then we need to show, in Eq. (\ref{etap}), that $X>Y$, i.e. 
\be
X - Y = P_{1}^{'} -P_{1} -P_{5}^{'} +
P_{5} +P_{2}^{'} -P_{2} +\frac{1}{2}(P_{3}^{'} -P_{3} )>0.
\label{eq64}
\ee
From the normalization condition, and  
the fact that $P_{4} > P_{4}^{'}$, we can write   
$P_{1} +P_{2} +P_{3} +P_{5} <P_{1}^{'} +P_{2}^{'} +P_{3}^{'} +P_{5}^{'}$,
or upon rearranging                                  
\be
P_{1}^{'} -P_{1} +P_{2}^{'} -P_{2} +P_{3}^{'} -P_{3} +P_{5}^{'} -P_{5} >0.
\label{in125}
\ee
As $P_{5} >P_{5}^{'}$,  
so if we substitute in Eq. (\ref{in125}), $P_{5} -P_{5}^{'} >0$ in place of 
$P_{5}^{'} -P_{5}$, the resulting expression is still positive. Thus
\be
P_{1}^{'} -P_{1} +P_{5} -P_{5}^{'} +P_{2}^{'} -P_{2} +P_{3}^{'} -P_{3} >0.
\label{125b}
\ee
Also, using the fact that $P_{3}^{'} -P_{3} < 0$, 
inequality (\ref{125b}) also implies:
\be
P_{1}^{'} -P_{1} + P_{5} - P_{5}^{'} +P_{2}^{'} -
P_{2} + \frac{1}{2}(P_{3}^{'} -P_{3} ) > 0,
\ee
which is Eq. (\ref{eq64}). 

\subsection{$P_{2}^{'} > P_{2} $}
\noindent Suppose we are in the regime of parameter values where 
$P_{2}^{'} >P_{2} $. We have earlier seen that the efficiency of 
the coupled model is higher than the uncoupled model, if $Y>0$.
Now, in $Y= (P_{2} -P_{2}^{'}) + (P_{3} -P_{3}^{'})/2 > 0$,
the first term is negative. Thus an essential condition for
$Y>0$ is that $P_{3} >P_{3}^{'}$. 
This inequality alongwith the following relations
\[P_{4} =\frac{e^{-4J/T_{1}}}{m} P_{3}, \quad    
P_{4}^{'}=\frac{e^{-4J/T_{2}}}{m} P_{3}^{'}, \]
implies that $P_{4} >P_{4}^{'}$. 
Similarly using the condition ${B_{2} }/{T_{2} } > {B_{1} }/{T_{1} }$, 
 we can show: $P_{5} >P_{5}^{'}$.  Thus given that 
 $P_{2}^{'} > P_{2}$ and the efficiency to be higher than 
 the uncoupled case, the following 
 relations hold:
\be
P_{3}^{} > P_{3}^{'},
\quad P_{4}^{} > P_{4}^{'}, \quad P_{5}^{} > P_{5}^{'}.
\label{pp3}
\ee 
Note that, it still leaves the relation between $P_1$ and $P_1'$
undetermined. However, this does not cause a difficulty 
and we can prove $X-Y >0$ here also, along the same lines
as shown above i.e. using
the normalization of probabilities and Eq. (\ref{pp3}). Thus 
 the same upper bound for efficiency, $\eta_{\rm ub}$, holds in this regime also. 
 %
%

\begin{thebibliography}{50}%
\makeatletter
\providecommand \@ifxundefined [1]{%
 \@ifx{#1\undefined}
}%
\providecommand \@ifnum [1]{%
 \ifnum #1\expandafter \@firstoftwo
 \else \expandafter \@secondoftwo
 \fi
}%
\providecommand \@ifx [1]{%
 \ifx #1\expandafter \@firstoftwo
 \else \expandafter \@secondoftwo
 \fi
}%
\providecommand \natexlab [1]{#1}%
\providecommand \enquote  [1]{``#1''}%
\providecommand \bibnamefont  [1]{#1}%
\providecommand \bibfnamefont [1]{#1}%
\providecommand \citenamefont [1]{#1}%
\providecommand \href@noop [0]{\@secondoftwo}%
\providecommand \href [0]{\begingroup \@sanitize@url \@href}%
\providecommand \@href[1]{\@@startlink{#1}\@@href}%
\providecommand \@@href[1]{\endgroup#1\@@endlink}%
\providecommand \@sanitize@url [0]{\catcode `\\12\catcode `\$12\catcode
  `\&12\catcode `\#12\catcode `\^12\catcode `\_12\catcode `\%12\relax}%
\providecommand \@@startlink[1]{}%
\providecommand \@@endlink[0]{}%
\providecommand \url  [0]{\begingroup\@sanitize@url \@url }%
\providecommand \@url [1]{\endgroup\@href {#1}{\urlprefix }}%
\providecommand \urlprefix  [0]{URL }%
\providecommand \Eprint [0]{\href }%
\providecommand \doibase [0]{http://dx.doi.org/}%
\providecommand \selectlanguage [0]{\@gobble}%
\providecommand \bibinfo  [0]{\@secondoftwo}%
\providecommand \bibfield  [0]{\@secondoftwo}%
\providecommand \translation [1]{[#1]}%
\providecommand \BibitemOpen [0]{}%
\providecommand \bibitemStop [0]{}%
\providecommand \bibitemNoStop [0]{.\EOS\space}%
\providecommand \EOS [0]{\spacefactor3000\relax}%
\providecommand \BibitemShut  [1]{\csname bibitem#1\endcsname}%
\let\auto@bib@innerbib\@empty
\bibitem [{\citenamefont {Scovil}\ and\ \citenamefont
  {Schulz-DuBois}(1959)}]{scovil}%
  \BibitemOpen
  \bibfield  {author} {\bibinfo {author} {\bibfnamefont {H.~E.~D.}\
  \bibnamefont {Scovil}}\ and\ \bibinfo {author} {\bibfnamefont {E.~O.}\
  \bibnamefont {Schulz-DuBois}},\ }\href {\doibase 10.1103/PhysRevLett.2.262}
  {\bibfield  {journal} {\bibinfo  {journal} {Phys. Rev. Lett.}\ }\textbf
  {\bibinfo {volume} {2}},\ \bibinfo {pages} {262} (\bibinfo {year}
  {1959})}\BibitemShut {NoStop}%
\bibitem [{\citenamefont {Quan}\ \emph {et~al.}(2007)\citenamefont {Quan},
  \citenamefont {Liu}, \citenamefont {Sun},\ and\ \citenamefont
  {Nori}}]{quan2006}%
  \BibitemOpen
  \bibfield  {author} {\bibinfo {author} {\bibfnamefont {H.~T.}\ \bibnamefont
  {Quan}}, \bibinfo {author} {\bibfnamefont {Y.-x.}\ \bibnamefont {Liu}},
  \bibinfo {author} {\bibfnamefont {C.~P.}\ \bibnamefont {Sun}}, \ and\
  \bibinfo {author} {\bibfnamefont {F.}~\bibnamefont {Nori}},\ }\href {\doibase
  10.1103/PhysRevE.76.031105} {\bibfield  {journal} {\bibinfo  {journal} {Phys.
  Rev. E}\ }\textbf {\bibinfo {volume} {76}},\ \bibinfo {pages} {031105}
  (\bibinfo {year} {2007})}\BibitemShut {NoStop}%
\bibitem [{\citenamefont {Kieu}(2006)}]{kieu2006}%
  \BibitemOpen
  \bibfield  {author} {\bibinfo {author} {\bibfnamefont {T.~D.}\ \bibnamefont
  {Kieu}},\ }\href@noop {} {\bibfield  {journal} {\bibinfo  {journal} {The
  European Physical Journal D-Atomic, Molecular, Optical and Plasma Physics}\
  }\textbf {\bibinfo {volume} {39}},\ \bibinfo {pages} {115} (\bibinfo {year}
  {2006})}\BibitemShut {NoStop}%
\bibitem [{\citenamefont {Abe}\ and\ \citenamefont {Okuyama}(2011)}]{Abe2011}%
  \BibitemOpen
  \bibfield  {author} {\bibinfo {author} {\bibfnamefont {S.}~\bibnamefont
  {Abe}}\ and\ \bibinfo {author} {\bibfnamefont {S.}~\bibnamefont {Okuyama}},\
  }\href {\doibase 10.1103/PhysRevE.83.021121} {\bibfield  {journal} {\bibinfo
  {journal} {Phys. Rev. E}\ }\textbf {\bibinfo {volume} {83}},\ \bibinfo
  {pages} {021121} (\bibinfo {year} {2011})}\BibitemShut {NoStop}%
%
  \bibitem [{\citenamefont {Allahverdyan}\ \emph {et~al.}(2008)\citenamefont {Allahverdyan},
   \citenamefont {Johal},\ and\ \citenamefont
  {Mahler}}]{Johal2008}%
  \BibitemOpen
  \bibfield  {author} {\bibinfo {author} {\bibfnamefont {A.~E.}\ \bibnamefont
  {Allahverdyan}}, \bibinfo {author} {\bibfnamefont {R.~S.}\ \bibnamefont {Johal}},
   \ and\ \bibinfo {author} {\bibfnamefont {G.}~\bibnamefont {Mahler}},\ }\href {\doibase
  10.1103/PhysRevE.77.041118} {\bibfield  {journal} {\bibinfo  {journal} {Phys.
  Rev. E}\ }\textbf {\bibinfo {volume} {78}},\ \bibinfo {pages} {041118}
  (\bibinfo {year} {2008})}\BibitemShut {NoStop}%
  %
  \bibitem [{\citenamefont {Uzdin}\ and\ \citenamefont
  {Kosloff}(2014)}]{uzdin2014}%
  \BibitemOpen
  \bibfield  {author} {\bibinfo {author} {\bibfnamefont {R.}~\bibnamefont
  {Uzdin}}\ and\ \bibinfo {author} {\bibfnamefont {R.}~\bibnamefont
  {Kosloff}},\ }\href@noop {} {\bibfield  {journal} {\bibinfo  {journal} {EPL
  (Europhysics Letters)}\ }\textbf {\bibinfo {volume} {108}},\ \bibinfo {pages}
  {40001} (\bibinfo {year} {2014})}\BibitemShut {NoStop}%
\bibitem [{\citenamefont {Thomas}\ and\ \citenamefont {Johal}(2011)}]{george}%
  \BibitemOpen
  \bibfield  {author} {\bibinfo {author} {\bibfnamefont {G.}~\bibnamefont
  {Thomas}}\ and\ \bibinfo {author} {\bibfnamefont {R.~S.}\ \bibnamefont
  {Johal}},\ }\href {\doibase 10.1103/PhysRevE.83.031135} {\bibfield  {journal}
  {\bibinfo  {journal} {Phys. Rev. E}\ }\textbf {\bibinfo {volume} {83}},\
  \bibinfo {pages} {031135} (\bibinfo {year} {2011})}\BibitemShut {NoStop}%
\bibitem [{\citenamefont {Ivanchenko}(2015)}]{ivanchenko2015}%
  \BibitemOpen
  \bibfield  {author} {\bibinfo {author} {\bibfnamefont {E.~A.}\ \bibnamefont
  {Ivanchenko}},\ }\href@noop {} {\bibfield  {journal} {\bibinfo  {journal}
  {Physical Review E}\ }\textbf {\bibinfo {volume} {92}},\ \bibinfo {pages}
  {032124} (\bibinfo {year} {2015})}\BibitemShut {NoStop}%
\bibitem [{\citenamefont {Altintas}\ and\ \citenamefont
  {M{\"u}stecapl{\i}o{\u{g}}lu}(2015)}]{altintas2015}%
  \BibitemOpen
  \bibfield  {author} {\bibinfo {author} {\bibfnamefont {F.}~\bibnamefont
  {Altintas}}\ and\ \bibinfo {author} {\bibfnamefont {{\"O}.~E.}\ \bibnamefont
  {M{\"u}stecapl{\i}o{\u{g}}lu}},\ }\href@noop {} {\bibfield  {journal}
  {\bibinfo  {journal} {Phys. Rev. E}\ }\textbf {\bibinfo {volume} {92}},\
  \bibinfo {pages} {022142} (\bibinfo {year} {2015})}\BibitemShut {NoStop}%
\bibitem [{\citenamefont {Thomas}\ \emph {et~al.}(2016)\citenamefont {Thomas},
  \citenamefont {Banik},\ and\ \citenamefont {Ghosh}}]{thomas2016}%
  \BibitemOpen
  \bibfield  {author} {\bibinfo {author} {\bibfnamefont {G.}~\bibnamefont
  {Thomas}}, \bibinfo {author} {\bibfnamefont {M.}~\bibnamefont {Banik}}, \
  and\ \bibinfo {author} {\bibfnamefont {S.}~\bibnamefont {Ghosh}},\
  }\href@noop {} {\bibfield  {journal} {\bibinfo  {journal} {arXiv preprint
  arXiv:1607.00994}\ } (\bibinfo {year} {2016})}\BibitemShut {NoStop}%
\bibitem [{\citenamefont {Rezek}\ and\ \citenamefont
  {Kosloff}(2006)}]{rezek2006}%
  \BibitemOpen
  \bibfield  {author} {\bibinfo {author} {\bibfnamefont {Y.}~\bibnamefont
  {Rezek}}\ and\ \bibinfo {author} {\bibfnamefont {R.}~\bibnamefont
  {Kosloff}},\ }\href@noop {} {\bibfield  {journal} {\bibinfo  {journal} {New
  Journal of Physics}\ }\textbf {\bibinfo {volume} {8}},\ \bibinfo {pages} {83}
  (\bibinfo {year} {2006})}\BibitemShut {NoStop}%
\bibitem [{\citenamefont {Ro{\ss}nagel}\ \emph {et~al.}(2014)\citenamefont
  {Ro{\ss}nagel}, \citenamefont {Abah}, \citenamefont {Schmidt-Kaler},
  \citenamefont {Singer},\ and\ \citenamefont {Lutz}}]{abah}%
  \BibitemOpen
  \bibfield  {author} {\bibinfo {author} {\bibfnamefont {J.}~\bibnamefont
  {Ro{\ss}nagel}}, \bibinfo {author} {\bibfnamefont {O.}~\bibnamefont {Abah}},
  \bibinfo {author} {\bibfnamefont {F.}~\bibnamefont {Schmidt-Kaler}}, \bibinfo
  {author} {\bibfnamefont {K.}~\bibnamefont {Singer}}, \ and\ \bibinfo {author}
  {\bibfnamefont {E.}~\bibnamefont {Lutz}},\ }\href@noop {} {\bibfield
  {journal} {\bibinfo  {journal} {Phys. Rev. Lett.}\ }\textbf {\bibinfo
  {volume} {112}},\ \bibinfo {pages} {030602} (\bibinfo {year}
  {2014})}\BibitemShut {NoStop}%
\bibitem [{\citenamefont {Dillenschneider}\ and\ \citenamefont
  {Lutz}(2009)}]{Lutz2009}%
  \BibitemOpen
  \bibfield  {author} {\bibinfo {author} {\bibfnamefont {R.}~\bibnamefont
  {Dillenschneider}}\ and\ \bibinfo {author} {\bibfnamefont {E.}~\bibnamefont
  {Lutz}},\ }\href@noop {} {\bibfield  {journal} {\bibinfo  {journal} {EPL
  (Europhysics Letters)}\ }\textbf {\bibinfo {volume} {88}},\ \bibinfo {pages}
  {50003} (\bibinfo {year} {2009})}\BibitemShut {NoStop}%
\bibitem [{\citenamefont {Agarwal}\ and\ \citenamefont
  {Chaturvedi}(2013)}]{agarwal2013}%
  \BibitemOpen
  \bibfield  {author} {\bibinfo {author} {\bibfnamefont {G.~S.}\ \bibnamefont
  {Agarwal}}\ and\ \bibinfo {author} {\bibfnamefont {S.}~\bibnamefont
  {Chaturvedi}},\ }\href@noop {} {\bibfield  {journal} {\bibinfo  {journal}
  {Phys. Rev. E}\ }\textbf {\bibinfo {volume} {88}},\ \bibinfo {pages} {012130}
  (\bibinfo {year} {2013})}\BibitemShut {NoStop}%
\bibitem [{\citenamefont {Oppenheim}\ \emph {et~al.}(2002)\citenamefont
  {Oppenheim}, \citenamefont {Horodecki}, \citenamefont {Horodecki},\ and\
  \citenamefont {Horodecki}}]{Horodecki2002}%
  \BibitemOpen
  \bibfield  {author} {\bibinfo {author} {\bibfnamefont {J.}~\bibnamefont
  {Oppenheim}}, \bibinfo {author} {\bibfnamefont {M.}~\bibnamefont
  {Horodecki}}, \bibinfo {author} {\bibfnamefont {P.}~\bibnamefont
  {Horodecki}}, \ and\ \bibinfo {author} {\bibfnamefont {R.}~\bibnamefont
  {Horodecki}},\ }\href {\doibase 10.1103/PhysRevLett.89.180402} {\bibfield
  {journal} {\bibinfo  {journal} {Phys. Rev. Lett.}\ }\textbf {\bibinfo
  {volume} {89}},\ \bibinfo {pages} {180402} (\bibinfo {year}
  {2002})}\BibitemShut {NoStop}%
\bibitem [{\citenamefont {Vedral}\ and\ \citenamefont
  {Kashefi}(2002)}]{Vedral2002}%
  \BibitemOpen
  \bibfield  {author} {\bibinfo {author} {\bibfnamefont {V.}~\bibnamefont
  {Vedral}}\ and\ \bibinfo {author} {\bibfnamefont {E.}~\bibnamefont
  {Kashefi}},\ }\href {\doibase 10.1103/PhysRevLett.89.037903} {\bibfield
  {journal} {\bibinfo  {journal} {Phys. Rev. Lett.}\ }\textbf {\bibinfo
  {volume} {89}},\ \bibinfo {pages} {037903} (\bibinfo {year}
  {2002})}\BibitemShut {NoStop}%
\bibitem [{\citenamefont {Kosloff}(2013)}]{Kosloff2013}%
  \BibitemOpen
  \bibfield  {author} {\bibinfo {author} {\bibfnamefont {R.}~\bibnamefont
  {Kosloff}},\ }\href {http://www.mdpi.com/1099-4300/15/6/2100} {\bibfield
  {journal} {\bibinfo  {journal} {Entropy}\ }\textbf {\bibinfo {volume} {15}},\
  \bibinfo {pages} {2100} (\bibinfo {year} {2013})}\BibitemShut {NoStop}%
\bibitem [{\citenamefont {Brand\~ao}\ \emph {et~al.}(2013)\citenamefont
  {Brand\~ao}, \citenamefont {Horodecki}, \citenamefont {Oppenheim},
  \citenamefont {Renes},\ and\ \citenamefont {Spekkens}}]{Spekkens2013}%
  \BibitemOpen
  \bibfield  {author} {\bibinfo {author} {\bibfnamefont {F.~G. S.~L.}\
  \bibnamefont {Brand\~ao}}, \bibinfo {author} {\bibfnamefont {M.}~\bibnamefont
  {Horodecki}}, \bibinfo {author} {\bibfnamefont {J.}~\bibnamefont
  {Oppenheim}}, \bibinfo {author} {\bibfnamefont {J.~M.}\ \bibnamefont
  {Renes}}, \ and\ \bibinfo {author} {\bibfnamefont {R.~W.}\ \bibnamefont
  {Spekkens}},\ }\href {\doibase 10.1103/PhysRevLett.111.250404} {\bibfield
  {journal} {\bibinfo  {journal} {Phys. Rev. Lett.}\ }\textbf {\bibinfo
  {volume} {111}},\ \bibinfo {pages} {250404} (\bibinfo {year}
  {2013})}\BibitemShut {NoStop}%
\bibitem [{\citenamefont {Gelbwaser-Klimovsky}\ \emph
  {et~al.}(2015)\citenamefont {Gelbwaser-Klimovsky}, \citenamefont {Niedenzu},\
  and\ \citenamefont {Kurizki}}]{Kurizki2015}%
  \BibitemOpen
  \bibfield  {author} {\bibinfo {author} {\bibfnamefont {D.}~\bibnamefont
  {Gelbwaser-Klimovsky}}, \bibinfo {author} {\bibfnamefont {W.}~\bibnamefont
  {Niedenzu}}, \ and\ \bibinfo {author} {\bibfnamefont {G.}~\bibnamefont
  {Kurizki}}\ }(\bibinfo  {publisher} {Academic Press},\ \bibinfo {year}
  {2015})\ pp.\ \bibinfo {pages} {329 -- 407}\BibitemShut {NoStop}%
\bibitem [{\citenamefont {Zhang}\ \emph {et~al.}(2007)\citenamefont {Zhang},
  \citenamefont {Liu}, \citenamefont {Chen},\ and\ \citenamefont
  {Li}}]{zhang2007}%
  \BibitemOpen
  \bibfield  {author} {\bibinfo {author} {\bibfnamefont {T.}~\bibnamefont
  {Zhang}}, \bibinfo {author} {\bibfnamefont {W.-T.}\ \bibnamefont {Liu}},
  \bibinfo {author} {\bibfnamefont {P.-X.}\ \bibnamefont {Chen}}, \ and\
  \bibinfo {author} {\bibfnamefont {C.-Z.}\ \bibnamefont {Li}},\ }\href@noop {}
  {\bibfield  {journal} {\bibinfo  {journal} {Phys. Rev. A}\ }\textbf {\bibinfo
  {volume} {75}},\ \bibinfo {pages} {062102} (\bibinfo {year}
  {2007})}\BibitemShut {NoStop}%
\bibitem [{\citenamefont {Wang}\ \emph {et~al.}(2009)\citenamefont {Wang},
  \citenamefont {Liu},\ and\ \citenamefont {He}}]{wang2009}%
  \BibitemOpen
  \bibfield  {author} {\bibinfo {author} {\bibfnamefont {H.}~\bibnamefont
  {Wang}}, \bibinfo {author} {\bibfnamefont {S.}~\bibnamefont {Liu}}, \ and\
  \bibinfo {author} {\bibfnamefont {J.}~\bibnamefont {He}},\ }\href@noop {}
  {\bibfield  {journal} {\bibinfo  {journal} {Phys. Rev. E}\ }\textbf {\bibinfo
  {volume} {79}},\ \bibinfo {pages} {041113} (\bibinfo {year}
  {2009})}\BibitemShut {NoStop}%
\bibitem [{\citenamefont {Abah}\ \emph {et~al.}(2012)\citenamefont {Abah},
  \citenamefont {Ro\ss{}nagel}, \citenamefont {Jacob}, \citenamefont {Deffner},
  \citenamefont {Schmidt-Kaler}, \citenamefont {Singer},\ and\ \citenamefont
  {Lutz}}]{Lutz2012}%
  \BibitemOpen
  \bibfield  {author} {\bibinfo {author} {\bibfnamefont {O.}~\bibnamefont
  {Abah}}, \bibinfo {author} {\bibfnamefont {J.}~\bibnamefont {Ro\ss{}nagel}},
  \bibinfo {author} {\bibfnamefont {G.}~\bibnamefont {Jacob}}, \bibinfo
  {author} {\bibfnamefont {S.}~\bibnamefont {Deffner}}, \bibinfo {author}
  {\bibfnamefont {F.}~\bibnamefont {Schmidt-Kaler}}, \bibinfo {author}
  {\bibfnamefont {K.}~\bibnamefont {Singer}}, \ and\ \bibinfo {author}
  {\bibfnamefont {E.}~\bibnamefont {Lutz}},\ }\href {\doibase
  10.1103/PhysRevLett.109.203006} {\bibfield  {journal} {\bibinfo  {journal}
  {Phys. Rev. Lett.}\ }\textbf {\bibinfo {volume} {109}},\ \bibinfo {pages}
  {203006} (\bibinfo {year} {2012})}\BibitemShut {NoStop}%
\bibitem [{\citenamefont {Li}\ \emph {et~al.}(2013)\citenamefont {Li},
  \citenamefont {Zou}, \citenamefont {Yu}, \citenamefont {Li}, \citenamefont
  {Xu},\ and\ \citenamefont {Shao}}]{Li2013}%
  \BibitemOpen
  \bibfield  {author} {\bibinfo {author} {\bibfnamefont {H.}~\bibnamefont
  {Li}}, \bibinfo {author} {\bibfnamefont {J.}~\bibnamefont {Zou}}, \bibinfo
  {author} {\bibfnamefont {W.-L.}\ \bibnamefont {Yu}}, \bibinfo {author}
  {\bibfnamefont {L.}~\bibnamefont {Li}}, \bibinfo {author} {\bibfnamefont
  {B.-M.}\ \bibnamefont {Xu}}, \ and\ \bibinfo {author} {\bibfnamefont
  {B.}~\bibnamefont {Shao}},\ }\href@noop {} {\bibfield  {journal} {\bibinfo
  {journal} {The European Physical Journal D}\ }\textbf {\bibinfo {volume}
  {67}},\ \bibinfo {pages} {134} (\bibinfo {year} {2013})}\BibitemShut
  {NoStop}%
\bibitem [{\citenamefont {Thomas}\ and\ \citenamefont
  {Johal}(2014)}]{Thomas2014}%
  \BibitemOpen
  \bibfield  {author} {\bibinfo {author} {\bibfnamefont {G.}~\bibnamefont
  {Thomas}}\ and\ \bibinfo {author} {\bibfnamefont {R.~S.}\ \bibnamefont
  {Johal}},\ }\href {\doibase 10.1140/epjb/e2014-50231-1} {\bibfield  {journal}
  {\bibinfo  {journal} {The European Physical Journal B}\ }\textbf {\bibinfo
  {volume} {87}},\ \bibinfo {pages} {166} (\bibinfo {year} {2014})}\BibitemShut
  {NoStop}%
\bibitem [{\citenamefont {Zhang}\ \emph {et~al.}(2014)\citenamefont {Zhang},
  \citenamefont {Bariani},\ and\ \citenamefont {Meystre}}]{Meystre2014}%
  \BibitemOpen
  \bibfield  {author} {\bibinfo {author} {\bibfnamefont {K.}~\bibnamefont
  {Zhang}}, \bibinfo {author} {\bibfnamefont {F.}~\bibnamefont {Bariani}}, \
  and\ \bibinfo {author} {\bibfnamefont {P.}~\bibnamefont {Meystre}},\ }\href
  {\doibase 10.1103/PhysRevLett.112.150602} {\bibfield  {journal} {\bibinfo
  {journal} {Phys. Rev. Lett.}\ }\textbf {\bibinfo {volume} {112}},\ \bibinfo
  {pages} {150602} (\bibinfo {year} {2014})}\BibitemShut {NoStop}%
\bibitem [{\citenamefont {Stefanatos}(2014)}]{Stefanotas2014}%
  \BibitemOpen
  \bibfield  {author} {\bibinfo {author} {\bibfnamefont {D.}~\bibnamefont
  {Stefanatos}},\ }\href {\doibase 10.1103/PhysRevE.90.012119} {\bibfield
  {journal} {\bibinfo  {journal} {Phys. Rev. E}\ }\textbf {\bibinfo {volume}
  {90}},\ \bibinfo {pages} {012119} (\bibinfo {year} {2014})}\BibitemShut
  {NoStop}%
\bibitem [{\citenamefont {Wu}\ \emph {et~al.}(2014)\citenamefont {Wu},
  \citenamefont {He}, \citenamefont {Ma},\ and\ \citenamefont {Wang}}]{Wu2014}%
  \BibitemOpen
  \bibfield  {author} {\bibinfo {author} {\bibfnamefont {F.}~\bibnamefont
  {Wu}}, \bibinfo {author} {\bibfnamefont {J.}~\bibnamefont {He}}, \bibinfo
  {author} {\bibfnamefont {Y.}~\bibnamefont {Ma}}, \ and\ \bibinfo {author}
  {\bibfnamefont {J.}~\bibnamefont {Wang}},\ }\href {\doibase
  10.1103/PhysRevE.90.062134} {\bibfield  {journal} {\bibinfo  {journal} {Phys.
  Rev. E}\ }\textbf {\bibinfo {volume} {90}},\ \bibinfo {pages} {062134}
  (\bibinfo {year} {2014})}\BibitemShut {NoStop}%
\bibitem [{\citenamefont {H{\"u}bner}\ \emph {et~al.}(2014)\citenamefont
  {H{\"u}bner}, \citenamefont {Lefkidis}, \citenamefont {Dong}, \citenamefont
  {Chaudhuri}, \citenamefont {Chotorlishvili},\ and\ \citenamefont
  {Berakdar}}]{hubner2014}%
  \BibitemOpen
  \bibfield  {author} {\bibinfo {author} {\bibfnamefont {W.}~\bibnamefont
  {H{\"u}bner}}, \bibinfo {author} {\bibfnamefont {G.}~\bibnamefont
  {Lefkidis}}, \bibinfo {author} {\bibfnamefont {C.~D.}\ \bibnamefont {Dong}},
  \bibinfo {author} {\bibfnamefont {D.}~\bibnamefont {Chaudhuri}}, \bibinfo
  {author} {\bibfnamefont {L.}~\bibnamefont {Chotorlishvili}}, \ and\ \bibinfo
  {author} {\bibfnamefont {J.}~\bibnamefont {Berakdar}},\ }\href@noop {}
  {\bibfield  {journal} {\bibinfo  {journal} {Phys. Rev. B}\ }\textbf {\bibinfo
  {volume} {90}},\ \bibinfo {pages} {024401} (\bibinfo {year}
  {2014})}\BibitemShut {NoStop}%
\bibitem [{\citenamefont {Pe\~na}\ and\ \citenamefont
  {Mu\~noz}(2015)}]{Pena2015}%
  \BibitemOpen
  \bibfield  {author} {\bibinfo {author} {\bibfnamefont {F.~J.}\ \bibnamefont
  {Pe\~na}}\ and\ \bibinfo {author} {\bibfnamefont {E.}~\bibnamefont
  {Mu\~noz}},\ }\href {\doibase 10.1103/PhysRevE.91.052152} {\bibfield
  {journal} {\bibinfo  {journal} {Phys. Rev. E}\ }\textbf {\bibinfo {volume}
  {91}},\ \bibinfo {pages} {052152} (\bibinfo {year} {2015})}\BibitemShut
  {NoStop}%
\bibitem [{\citenamefont {Alecce}\ \emph {et~al.}(2015)\citenamefont {Alecce},
  \citenamefont {Galve}, \citenamefont {Gullo}, \citenamefont {Dell’Anna},
  \citenamefont {Plastina},\ and\ \citenamefont {Zambrini}}]{Zambrini2015}%
  \BibitemOpen
  \bibfield  {author} {\bibinfo {author} {\bibfnamefont {A.}~\bibnamefont
  {Alecce}}, \bibinfo {author} {\bibfnamefont {F.}~\bibnamefont {Galve}},
  \bibinfo {author} {\bibfnamefont {N.~L.}\ \bibnamefont {Gullo}}, \bibinfo
  {author} {\bibfnamefont {L.}~\bibnamefont {Dell’Anna}}, \bibinfo {author}
  {\bibfnamefont {F.}~\bibnamefont {Plastina}}, \ and\ \bibinfo {author}
  {\bibfnamefont {R.}~\bibnamefont {Zambrini}},\ }\href
  {http://stacks.iop.org/1367-2630/17/i=7/a=075007} {\bibfield  {journal}
  {\bibinfo  {journal} {New Journal of Physics}\ }\textbf {\bibinfo {volume}
  {17}},\ \bibinfo {pages} {075007} (\bibinfo {year} {2015})}\BibitemShut
  {NoStop}%
\bibitem [{\citenamefont {Wang}\ \emph {et~al.}(2015)\citenamefont {Wang},
  \citenamefont {Ye}, \citenamefont {Lai}, \citenamefont {Li},\ and\
  \citenamefont {He}}]{Jizhou2015}%
  \BibitemOpen
  \bibfield  {author} {\bibinfo {author} {\bibfnamefont {J.}~\bibnamefont
  {Wang}}, \bibinfo {author} {\bibfnamefont {Z.}~\bibnamefont {Ye}}, \bibinfo
  {author} {\bibfnamefont {Y.}~\bibnamefont {Lai}}, \bibinfo {author}
  {\bibfnamefont {W.}~\bibnamefont {Li}}, \ and\ \bibinfo {author}
  {\bibfnamefont {J.}~\bibnamefont {He}},\ }\href {\doibase
  10.1103/PhysRevE.91.062134} {\bibfield  {journal} {\bibinfo  {journal} {Phys.
  Rev. E}\ }\textbf {\bibinfo {volume} {91}},\ \bibinfo {pages} {062134}
  (\bibinfo {year} {2015})}\BibitemShut {NoStop}%
\bibitem [{\citenamefont {Long}\ and\ \citenamefont {Liu}(2015)}]{Long2015}%
  \BibitemOpen
  \bibfield  {author} {\bibinfo {author} {\bibfnamefont {R.}~\bibnamefont
  {Long}}\ and\ \bibinfo {author} {\bibfnamefont {W.}~\bibnamefont {Liu}},\
  }\href {\doibase 10.1103/PhysRevE.91.062137} {\bibfield  {journal} {\bibinfo
  {journal} {Phys. Rev. E}\ }\textbf {\bibinfo {volume} {91}},\ \bibinfo
  {pages} {062137} (\bibinfo {year} {2015})}\BibitemShut {NoStop}%
\bibitem [{\citenamefont {Zheng}\ and\ \citenamefont
  {Poletti}(2015)}]{Poletti2015}%
  \BibitemOpen
  \bibfield  {author} {\bibinfo {author} {\bibfnamefont {Y.}~\bibnamefont
  {Zheng}}\ and\ \bibinfo {author} {\bibfnamefont {D.}~\bibnamefont
  {Poletti}},\ }\href {\doibase 10.1103/PhysRevE.92.012110} {\bibfield
  {journal} {\bibinfo  {journal} {Phys. Rev. E}\ }\textbf {\bibinfo {volume}
  {92}},\ \bibinfo {pages} {012110} (\bibinfo {year} {2015})}\BibitemShut
  {NoStop}%
\bibitem [{\citenamefont {Jaramillo}\ \emph {et~al.}(2016)\citenamefont
  {Jaramillo}, \citenamefont {Beau},\ and\ \citenamefont {del
  Campo}}]{Campo2016new}%
  \BibitemOpen
  \bibfield  {author} {\bibinfo {author} {\bibfnamefont {J.}~\bibnamefont
  {Jaramillo}}, \bibinfo {author} {\bibfnamefont {M.}~\bibnamefont {Beau}}, \
  and\ \bibinfo {author} {\bibfnamefont {A.}~\bibnamefont {del Campo}},\
  }\href@noop {} {\bibfield  {journal} {\bibinfo  {journal} {New J. Phys.}\
  }\textbf {\bibinfo {volume} {18}},\ \bibinfo {pages} {075019} (\bibinfo
  {year} {2016})}\BibitemShut {NoStop}%
\bibitem [{\citenamefont {Beau}\ \emph {et~al.}(2016)\citenamefont {Beau},
  \citenamefont {Jaramillo},\ and\ \citenamefont {del Campo}}]{Campo2016}%
  \BibitemOpen
  \bibfield  {author} {\bibinfo {author} {\bibfnamefont {M.}~\bibnamefont
  {Beau}}, \bibinfo {author} {\bibfnamefont {J.}~\bibnamefont {Jaramillo}}, \
  and\ \bibinfo {author} {\bibfnamefont {A.}~\bibnamefont {del Campo}},\
  }\href@noop {} {\bibfield  {journal} {\bibinfo  {journal} {Entropy}\ }\textbf
  {\bibinfo {volume} {18}},\ \bibinfo {pages} {168} (\bibinfo {year}
  {2016})}\BibitemShut {NoStop}%
\bibitem [{\citenamefont {{\c{C}}akmak}\ \emph {et~al.}(2016)\citenamefont
  {{\c{C}}akmak}, \citenamefont {Altintas},\ and\ \citenamefont
  {E.~M{\"u}stecapl{\i}o{\u{g}}lu}}]{Ferdi2016}%
  \BibitemOpen
  \bibfield  {author} {\bibinfo {author} {\bibfnamefont {S.}~\bibnamefont
  {{\c{C}}akmak}}, \bibinfo {author} {\bibfnamefont {F.}~\bibnamefont
  {Altintas}}, \ and\ \bibinfo {author} {\bibfnamefont {{\"O}.}~\bibnamefont
  {E.~M{\"u}stecapl{\i}o{\u{g}}lu}},\ }\href {\doibase
  10.1140/epjp/i2016-16197-0} {\bibfield  {journal} {\bibinfo  {journal} {The
  European Physical Journal Plus}\ }\textbf {\bibinfo {volume} {131}},\
  \bibinfo {pages} {197} (\bibinfo {year} {2016})}\BibitemShut {NoStop}%
\bibitem [{\citenamefont {Leggio}\ and\ \citenamefont
  {Antezza}(2016)}]{Leggio2016}%
  \BibitemOpen
  \bibfield  {author} {\bibinfo {author} {\bibfnamefont {B.}~\bibnamefont
  {Leggio}}\ and\ \bibinfo {author} {\bibfnamefont {M.}~\bibnamefont
  {Antezza}},\ }\href {\doibase 10.1103/PhysRevE.93.022122} {\bibfield
  {journal} {\bibinfo  {journal} {Phys. Rev. E}\ }\textbf {\bibinfo {volume}
  {93}},\ \bibinfo {pages} {022122} (\bibinfo {year} {2016})}\BibitemShut
  {NoStop}%
\bibitem [{\citenamefont {Manzano}\ \emph {et~al.}(2016)\citenamefont
  {Manzano}, \citenamefont {Galve}, \citenamefont {Zambrini},\ and\
  \citenamefont {Parrondo}}]{Manzano2016}%
  \BibitemOpen
  \bibfield  {author} {\bibinfo {author} {\bibfnamefont {G.}~\bibnamefont
  {Manzano}}, \bibinfo {author} {\bibfnamefont {F.}~\bibnamefont {Galve}},
  \bibinfo {author} {\bibfnamefont {R.}~\bibnamefont {Zambrini}}, \ and\
  \bibinfo {author} {\bibfnamefont {J.~M.~R.}\ \bibnamefont {Parrondo}},\
  }\href {\doibase 10.1103/PhysRevE.93.052120} {\bibfield  {journal} {\bibinfo
  {journal} {Phys. Rev. E}\ }\textbf {\bibinfo {volume} {93}},\ \bibinfo
  {pages} {052120} (\bibinfo {year} {2016})}\BibitemShut {NoStop}%
\bibitem [{\citenamefont {Insinga}\ \emph {et~al.}(2016)\citenamefont
  {Insinga}, \citenamefont {Andresen},\ and\ \citenamefont
  {Salamon}}]{Salamon2016}%
  \BibitemOpen
  \bibfield  {author} {\bibinfo {author} {\bibfnamefont {A.}~\bibnamefont
  {Insinga}}, \bibinfo {author} {\bibfnamefont {B.}~\bibnamefont {Andresen}}, \
  and\ \bibinfo {author} {\bibfnamefont {P.}~\bibnamefont {Salamon}},\ }\href
  {\doibase 10.1103/PhysRevE.94.012119} {\bibfield  {journal} {\bibinfo
  {journal} {Phys. Rev. E}\ }\textbf {\bibinfo {volume} {94}},\ \bibinfo
  {pages} {012119} (\bibinfo {year} {2016})}\BibitemShut {NoStop}%
\bibitem [{\citenamefont {Zheng}\ \emph {et~al.}(2016)\citenamefont {Zheng},
  \citenamefont {H\"anggi},\ and\ \citenamefont {Poletti}}]{Poletti2016}%
  \BibitemOpen
  \bibfield  {author} {\bibinfo {author} {\bibfnamefont {Y.}~\bibnamefont
  {Zheng}}, \bibinfo {author} {\bibfnamefont {P.}~\bibnamefont {H\"anggi}}, \
  and\ \bibinfo {author} {\bibfnamefont {D.}~\bibnamefont {Poletti}},\ }\href
  {\doibase 10.1103/PhysRevE.94.012137} {\bibfield  {journal} {\bibinfo
  {journal} {Phys. Rev. E}\ }\textbf {\bibinfo {volume} {94}},\ \bibinfo
  {pages} {012137} (\bibinfo {year} {2016})}\BibitemShut {NoStop}%
\bibitem [{\citenamefont {Karimi}\ and\ \citenamefont
  {Pekola}(2016)}]{Pekola2016}%
  \BibitemOpen
  \bibfield  {author} {\bibinfo {author} {\bibfnamefont {B.}~\bibnamefont
  {Karimi}}\ and\ \bibinfo {author} {\bibfnamefont {J.~P.}\ \bibnamefont
  {Pekola}},\ }\href {\doibase 10.1103/PhysRevB.94.184503} {\bibfield
  {journal} {\bibinfo  {journal} {Phys. Rev. B}\ }\textbf {\bibinfo {volume}
  {94}},\ \bibinfo {pages} {184503} (\bibinfo {year} {2016})}\BibitemShut
  {NoStop}%
\bibitem [{\citenamefont {Chand}\ and\ \citenamefont
  {Biswas}(2017)}]{Biswas2017}%
  \BibitemOpen
  \bibfield  {author} {\bibinfo {author} {\bibfnamefont {S.}~\bibnamefont
  {Chand}}\ and\ \bibinfo {author} {\bibfnamefont {A.}~\bibnamefont {Biswas}},\
  }\href {\doibase 10.1103/PhysRevE.95.032111} {\bibfield  {journal} {\bibinfo
  {journal} {Phys. Rev. E}\ }\textbf {\bibinfo {volume} {95}},\ \bibinfo
  {pages} {032111} (\bibinfo {year} {2017})}\BibitemShut {NoStop}%
\bibitem [{\citenamefont {Newman}\ \emph {et~al.}(2017)\citenamefont {Newman},
  \citenamefont {Mintert},\ and\ \citenamefont {Nazir}}]{Newman2017}%
  \BibitemOpen
  \bibfield  {author} {\bibinfo {author} {\bibfnamefont {D.}~\bibnamefont
  {Newman}}, \bibinfo {author} {\bibfnamefont {F.}~\bibnamefont {Mintert}}, \
  and\ \bibinfo {author} {\bibfnamefont {A.}~\bibnamefont {Nazir}},\ }\href
  {\doibase 10.1103/PhysRevE.95.032139} {\bibfield  {journal} {\bibinfo
  {journal} {Phys. Rev. E}\ }\textbf {\bibinfo {volume} {95}},\ \bibinfo
  {pages} {032139} (\bibinfo {year} {2017})}\BibitemShut {NoStop}%
\bibitem [{\citenamefont {Huang}\ \emph {et~al.}(2013)\citenamefont {Huang},
  \citenamefont {Wang},\ and\ \citenamefont {Yi}}]{Huang2013}%
  \BibitemOpen
  \bibfield  {author} {\bibinfo {author} {\bibfnamefont {X.~L.}\ \bibnamefont
  {Huang}}, \bibinfo {author} {\bibfnamefont {L.~C.}\ \bibnamefont {Wang}}, \
  and\ \bibinfo {author} {\bibfnamefont {X.~X.}\ \bibnamefont {Yi}},\ }\href
  {\doibase 10.1103/PhysRevE.87.012144} {\bibfield  {journal} {\bibinfo
  {journal} {Phys. Rev. E}\ }\textbf {\bibinfo {volume} {87}},\ \bibinfo
  {pages} {012144} (\bibinfo {year} {2013})}\BibitemShut {NoStop}%
\bibitem [{\citenamefont {Huang}\ \emph {et~al.}(2014)\citenamefont {Huang},
  \citenamefont {Liu}, \citenamefont {Wang},\ and\ \citenamefont
  {Niu}}]{Huang2014}%
  \BibitemOpen
  \bibfield  {author} {\bibinfo {author} {\bibfnamefont {X.~L.}\ \bibnamefont
  {Huang}}, \bibinfo {author} {\bibfnamefont {Y.}~\bibnamefont {Liu}}, \bibinfo
  {author} {\bibfnamefont {Z.}~\bibnamefont {Wang}}, \ and\ \bibinfo {author}
  {\bibfnamefont {X.~Y.}\ \bibnamefont {Niu}},\ }\href {\doibase
  10.1140/epjp/i2014-14004-8} {\bibfield  {journal} {\bibinfo  {journal} {The
  European Physical Journal Plus}\ }\textbf {\bibinfo {volume} {129}},\
  \bibinfo {pages} {4} (\bibinfo {year} {2014})}\BibitemShut {NoStop}%
  \bibitem{Scullybook} M. O. Scully and M. S. Zubairy, {\it Quantum Optics},
  Cambridge University Press (1997).
  \bibitem{Kurizki2015n} 
    D. Gelbwaser-Klimovsky, W. Niedenzu, P. Brumer, and G. Kurizki,
    Sc. Rep. {\bf 5}, 14413 (2015).
  %
  \bibitem [{\citenamefont {Born}\ and\ \citenamefont {Fock}(1928)}]{born1928}%
  \BibitemOpen
  \bibfield  {author} {\bibinfo {author} {\bibfnamefont {M.}~\bibnamefont
  {Born}}\ and\ \bibinfo {author} {\bibfnamefont {V.}~\bibnamefont {Fock}},\
  }\href@noop {} {\bibfield  {journal} {\bibinfo  {journal} {Zeitschrift
  f{\"u}r Physik A Hadrons and Nuclei}\ }\textbf {\bibinfo {volume} {51}},\
  \bibinfo {pages} {165} (\bibinfo {year} {1928})}\BibitemShut {NoStop}%
\bibitem{Comment2} For a more general scenario, where the free hamiltonian
also contains a term such as $2B_0\left( s_{x}^{(1)} \otimes I + 
I \otimes  s_{x}^{(2)} \right)$, i.e. 
a constant field $B_0$ applied in, say,  x-direction 
during the cycle, we can have $\eta_{\rm loc} \ne \eta_0$.
  %
  \bibitem{Comment} It is possible that 
${\rm Tr}[H_1\Delta \rho ] < 0$ in Eq. (\ref{qh}), 
but still we have $Q_h >0$. This, of course, requires
${\rm Tr}[H_{\rm int} \Delta \rho ] >0$, but it implies
that locally the substance undergoes a refrigeration
cycle,
but globally, we have the operation of an engine.
An example, with two coupled spin-1/2 particles
and $B_2 > B_1$ was given in Ref. \cite{george}.
%
  \end{thebibliography}
\end{document}